\documentclass[11pt,onecolumn]{article} 
\usepackage{latex8}
\usepackage{times}

\usepackage{algorithm}
\usepackage{algorithmic}
\usepackage{graphicx}

\begin{document}

\title{GCD Computation of $n$ Integers}


\author{
 Shri Prakash Dwivedi \thanks{Email: shriprakashdwivedi@gbpuat-tech.ac.in}\\ 
}

\maketitle

\begin{abstract}
Greatest Common Divisor (GCD) computation is one of the most important operation of algorithmic number theory. In this paper we present the algorithms for GCD computation of $n$ integers. We extend the Euclid's algorithm and binary GCD algorithm to compute the GCD of more than two integers.
\end{abstract}

\section{Introduction}
Greatest Common Divisor (GCD) of two integers is the largest integer that divides both integers. GCD computation has applications in rational arithmetic for simplifying numerator and denominator of a rational number. Other applications of GCD includes integer factoring, modular arithmetic and random number generation.
Euclid's algorithm is one of the most important method to compute the GCD of two integers. Lehmer [5] proposed the improvement over Euclid's algorithm for large integers. Blankinship [2] described a new version of Euclidean algorithm. Stein [7] described the binary GCD algorithm which uses only division by 2 (considered as shift operation) and subtract operation instead of expensive multiplication and division operations. Asymptotic complexity of Euclid's, binary GCD and Lehmer's algorithms remains $O(n^{2})$ [4]. GCD of two integers $a$ and $b$ can be computed in $O((\log a) (\log b))$ bit operations [3]. Knuth and Schonhage proposed subquadratic algorithm for GCD computation. Stehle and Zimmermann [6] described binary recursive GCD algorithm. Sorenson proposed the generalization of the binary and left-shift binary algorithm [8]. Asymptotically fastest GCD algorithms have running time of $O(n \log^2 n \log\log n)$ bit operations [1].

\indent In this paper we describe the algorithms for GCD computation of $n$ integers. We extend the Euclid's algorithm to compute GCD of more than two integers as explained in [4]. We also extend the binary GCD algorithm so that it can be used to compute GCD of many integers \\
\indent This paper is organized as follows. Section II describes background and motivation. Section III presents GCD algorithms for $n$ integers and their features. 
Finally, section IV contains conclusion.

\section{Background and Motivation}
GCD of two integers $a, b$ can be defined formally as :
$$ GCD(a,b)= max\{m \mbox{ such that } m|a \mbox{ and } m|b \} $$
GCD of two positive integers $a, b$ such that $a>b$ can be computed using the following recursive method:
$$GCD(a, b)=GCD(b, a \bmod b) $$
GCD operation follows the associative property, $GCD(a, GCD(b, c))= GCD(GCD(a,b),c)$. Further we can show that\\\\ 
\textbf{Proposition 1}\\
$$GCD(a,b,c)= GCD(GCD(a,b),c) $$
 Let $d$ be the GCD of $a,b$ and $c$. then by definition
   $$ d|GCD(a,b,c) $$
   $$ \Rightarrow d|a,b,c $$
   $$ \Rightarrow d|a,b \textit{ and } c $$
   $$ \Rightarrow d|GCD(a,b),c $$
   $$ \Rightarrow d|GCD(GCD(a,b),c) $$
 Using induction the above Proposition can be extended to a list of $n$ integers.\\\\
 \textbf{Proposition 2}\\
$$GCD(a_1,a_2,a_3...,a_n)= GCD(a_1, GCD(a_2,a_3,...,a_n)) $$
Above expression can be used to compute GCD of more than two integers. But if the list of integers ($n$) are very large and the pair of integers for which GCD is computed in a particular step is not randomly selected then the computation can be expensive.\\
In this paper we discuss the algorithms for GCD computation of many integers by actually extending the properties of the GCD computation of two integers.

\section{GCD Algorithms}
\subsection{Extension to Euclid's Algorithm}
Euclid's algorithm for GCD of two integers can be extended to GCD of $n$ integers (in this paper integers are non-negative unless otherwise specified) using the following facts:\\
Let $a_1, a_2, ..., a_n$ be $n$ integers. \\\\
\textbf{Proposition 3}\\
$$ GCD(0,0,...,0)=0 $$
This can be established by taking the convention $GCD(0,0)$ for $n=2$ and extending it to $n$ numbers.\\\\
\textbf{Proposition 4}\\
$$ GCD(a_1,0,...,0)=a_1 $$
\textit{Proof}: It follows from the fact $GCD(a_1,0)=a_1$ together with associativity of GCD. \\
$GCD(a_1,0,...,0)$\\
$=GCD(a_1,GCD(0,0,...0))=GCD(a_1,0)=a_1$\\
or, $GCD(...GCD(GCD(a_1,0),0)...,0)=a_1$\\\\
\textbf{Proposition 5}\\
\textit{Let} $min(a_1,a_2,...,a_n)=a_1$ \textit{then}
$$ GCD(a_1, a_2, ..., a_n)=GCD(a_1, a_2 \bmod a_1,..., a_n \bmod a_1) $$
\textit{Proof}: Note that if $a_1<a_2$ then $GCD(a_1,a_2)=GCD(a_1,a_2-xa_1)$ for any integer $x$. As any common divisor $a_1$ and $a_2$ will be divisor of both $a_1$ and $a_2-xa_1$, and also any common divisor of $a_1$ and $a_2-xa_1$ will divide both $a_1$ and $a_2$.
Now, Proposition 5 follows for $n=2$. Similarly, it can be extended using induction for $n$ numbers.\\

GCD-N is described in Algorithm 1. In this algorithm if at any time there is only one non-zero $a_i$, it will be considered as GCD, in other cases all $a_i$'s will be reduced by $a_i \bmod a_j$, where $a_j$ is the least non-zero integer at a given iteration. Detailed description of GCD-N is as follows: first while loop is used to check that at least one $a_i$ is non zero. If all numbers are 0, final GCD will be taken as 0. First for loop is used to store least non-zero integer in $a_1$. Second for loop is used to reduce other $a_i$'s to $a_i \bmod a_1$ except $a_1$. Third for loop is used to store largest and second largest integers in $a_n$ and $a_{n-1}$ respectively. Finally, if loop is used to check whether $a_n$ is the only non-zero integer, if this is the case, $a_n$ is returned otherwise while loop is called again.

\begin{algorithm}
\caption{\bf :  GCD-N $(a_1,a_2,...,a_n)$}
\begin{algorithmic}
\STATE \textbf{INPUT}: $a_1,a_2,...,a_n$
\STATE \textbf{OUTPUT}: $GCD(a_1,a_2,...,a_n)$
   \WHILE { $(a_1$ or $a_2$ or ... or $a_n)$ } 
   {
   \FOR {$(i\leftarrow 2; i \leq n; i\leftarrow i+1)$ }
   {
   \IF {$(a_1 > a_i$ and $a_i \neq 0) $}
   \STATE $temp \leftarrow a_i $ \\
   \STATE $a_i \leftarrow a_1 $ \\
   \STATE $a_1 \leftarrow temp $ 
   \ENDIF
   }
   \ENDFOR

   \FOR {$(i\leftarrow 2; i \leq n; i\leftarrow i+1)$ }
   {
   \STATE $a_i \leftarrow a_i \bmod a_1$
   }
   \ENDFOR

   \FOR {$(i\leftarrow n-2; i \geq 1; i\leftarrow i-1)$ }
   {
   \IF {$(a_i > a_n)$}
   \STATE $temp \leftarrow a_i $ \\
   \STATE $a_i \leftarrow a_n $ \\
   \STATE $a_n \leftarrow temp $ \\
   \ENDIF
   \IF {$(a_i < a_n$ and $a_i > a_{n-1})$}
   \STATE $temp \leftarrow a_i $ \\
   \STATE $a_i \leftarrow a_{n-1} $ \\
   \STATE $a_{n-1} \leftarrow temp $ \\
   \ENDIF
   }
   \ENDFOR

   \IF {$(a_n \neq 0$ and $a_{n-1} == 0) $}
   \RETURN $a_n$
   \ENDIF
   }
   \ENDWHILE
\end{algorithmic}
\end{algorithm}

Correctness of GCD-N algorithm follows from the Proposition 3, 4 and 5. \\
Example 1: Let $a_1=22, a_2=36, a_3=74, a_4=98$ then \\
$GCD(a_1, a_2, a_3, a_4)$\\
$\Rightarrow GCD(22, 36, 74, 98)$\\
$\Rightarrow GCD(22, 36 \bmod 22, 74 \bmod 22, 98 \bmod 22)$\\
$\Rightarrow GCD(22, 14, 8, 10)$\\
$\Rightarrow GCD(8, 14, 22, 10)$\\
$\Rightarrow GCD(8, 14 \bmod 8, 22 \bmod 8, 10 \bmod 8)$\\
$\Rightarrow GCD(8, 6, 6, 2)$\\
$\Rightarrow GCD(2, 6, 6, 8)$\\
$\Rightarrow GCD(2, 6 \bmod 2, 6 \bmod 2, 8 \bmod 2)$\\
$\Rightarrow GCD(2, 0, 0, 0)$\\
$\Rightarrow 2 $

\subsection{Extension to Binary GCD Algorithm}
GCD of $n$ integers can be computed using only shift and subtract operation similar to binary GCD algorithm. This algorithm is based on the following facts.\\\\
\textbf{Proposition 6}\\
\textit{Let} $a_1, a_2, ..., a_n$ \textit{be even integers} \textit{then}\\
$$ GCD(a_1,a_2,...,a_n)=2.GCD(a_1/2,a_2/2,...,a_n/2) $$
\textit{Proof}: Consider the above statement for $n=2$. Let $a_1'=a_1/2$ and $a_2'=a_2/2$. By definition $GCD(a_1,a_2)$ is least positive value of $a_1m_1+a_2m_2$ where $m_1$ and $m_2$ range over all integers.
Now, $GCD(a_1,a_2)=GCD(2a_1',2a_2')$ \\
$\Rightarrow$ least positive value of $2a_1'm_1+2a_2'm_2$ \\
$\Rightarrow$ 2.\{least positive value of $a_1'm_1+a_2'm_2$\} \\
$\Rightarrow 2.GCD(a_1',a_2')$ \\
$\Rightarrow 2.GCD(a_1/2,a_2/2)$ \\
Therefore, above statement is proved for $n=2$. Now, using induction the statement can be shown to hold for $n$.\\
Alternatively one can use the general definition:
\begin{equation}
 GCD(a_1,a_2,...,a_n)=\prod_{p \mbox{ prime}} p^{min(a_{1_p},a_{2_p},...,a_{n_p})}
\end{equation}
Where each $a_i$ is expressed in its unique prime factorization in ascending order.
$$a_i = 2^{a_{i_2}}.3^{a_{i_3}}.5^{a_{i_5}}...=\prod_{p \mbox{ prime}}p^{a_{i_p}}$$
Now, using the Equation 1, Proposition 6 can be easily established.
Since $a_1, a_2, ..., a_n$ are all even integers. Let $a_1'=a_1/2,a_2'=a_2/2,...,a_n'=a_n/2$ then \\
$ GCD(a_1,a_2,...,a_n) $\\
$\Rightarrow GCD(2a_1',2a_2',...,2a_n')$\\
$\Rightarrow 2.GCD(a_1',a_2',...,a_n')$\\
$\Rightarrow 2.GCD(a_1/2,a_2/2,...,a_n/2)$\\\\
\textbf{Proposition 7}\\
\textit{Let} $a_1, a_2, ..., a_m$ \textit{be odd and} $a_{m+1},a_{m+2},...,a_n$ \textit{be even integers} \textit{then}\\
$$ GCD(a_1,a_2,...,a_m,...,a_n)=$$ $$GCD(a_1,...,a_m,a_{m+1}/2,...,a_n/2) $$
\textit{Proof}: Since $a_1, a_2, ..., a_m$ are odd integers, they are not divisible by 2. But $a_{m+1},a_{m+2},...,a_n$ are even, and hence divisible by 2. Let $a_{m+1}'=a_{m+1}/2,a_{m+2}'=a_{m+2}/2,...,a_n'=a_n/2$. Then using the Equation 1, again:\\
$ GCD(a_1,a_2,...,a_m,a_{m+1},a_{m+2}...,a_n)$\\
$\Rightarrow GCD(a_1,a_2,...,a_m,2a_{m+1}',2a_{m+2}'...,2a_n')$\\
$\Rightarrow GCD(a_1,a_2,...,a_m,a_{m+1}',a_{m+2}'...,a_n')$, Since 2 is not common factor  \\
$\Rightarrow GCD(a_1,a_2,...,a_m,a_{m+1}/2,a_{m+2}/2...,a_n/2)$\\\\
\textbf{Proposition 8}\\
\textit{Let} $a_1, a_2, ..., a_n$ \textit{be odd integers} \\
\textit{Let} $min(a_1,a_2,...,a_n)=a_1$ \textit{then}
$$ GCD(a_1, a_2, ..., a_n)=GCD(a_1, (a_2-a_1)/2,..., (a_n-a_1)/2) $$
\textit{Proof}: Using the explanation in Proposition 5, if $a_1<a_2$ then $GCD(a_1,a_2)=GCD(a_1,a_2-xa_1)$, by putting $x=1$ and extending it to the case of $n$ integers, we can write.\\
$ GCD(a_1, a_2, ..., a_n)$
$=GCD(a_1, (a_2-a_1),..., (a_n-a_1)) $\\
Now all terms $(a_2-a_1),(a_3-a_1),...,(a_n-a_1)$ are even except first term $a_1$ which is still odd. Therefore we can use the Proposition 7 to further reduce it.
$GCD(a_1, (a_2-a_1),..., (a_n-a_1)) $\\
$=GCD(a_1, (a_2-a_1)/2,..., (a_n-a_1)/2) $\\

BINARY-GCD-N is described in Algorithm 2. In this algorithm first while loop is used to check how many iterations all of the $n$ integers are divisible by 2, and this value is stored in counter $p$. Second while loop is used to check for all integers to be not zero. First for loop is used to store least non-zero integer in $a_1$. Second for loop is used to reduce other $a_i$'s to $a_i - a_1$ except $a_1$, and this process is repeated until all $a_i$'s except one are zero. Correctness of BINARY-GCD-N algorithm follows from Proposition 6,7 and 8.

\begin{algorithm}
\caption{\bf :  BINARY-GCD-N $(a_1,a_2,...,a_n)$}
\begin{algorithmic}
\STATE \textbf{INPUT}: $a_1,a_2,...,a_n$
\STATE \textbf{OUTPUT}: $GCD(a_1,a_2,...,a_n)$
   \WHILE { $(a_1 \bmod 2 ==0$ and $a_2 \bmod 2 ==0$ and ... and $a_n \bmod 2 ==0)$ } 
   {
   \STATE $a_1 \leftarrow a_1/2 $ \\
   \STATE $a_2 \leftarrow a_2/2 $ \\
   \STATE .......................
   \STATE $a_n \leftarrow a_n/2 $
   \STATE $p \leftarrow p+1 $
   }
   \ENDWHILE

   \WHILE { $(a_2$ or $a_3$ or ... or $a_n)$ } 
   {
     \WHILE {$(a_1 \bmod 2 ==0)$} \STATE $a_1 \leftarrow a_1/2 $ \\
     \ENDWHILE
     \WHILE {$(a_2 \bmod 2 ==0)$} \STATE $a_2 \leftarrow a_2/2 $ \\
     \ENDWHILE 
     \STATE .......................
     \WHILE {$(a_n \bmod 2 ==0)$} \STATE $a_n \leftarrow a_n/2 $ \\
     \ENDWHILE

     \FOR {$(i\leftarrow 2; i \leq n; i\leftarrow i+1)$ }
     {
      \IF {$(a_1 > a_i) $}
      \STATE $temp \leftarrow a_1 $ \\
      \STATE $a_1 \leftarrow a_i $ \\
      \STATE $a_i \leftarrow temp $ 
      \ENDIF
     }
     \ENDFOR
     \FOR {$(i\leftarrow 2; i \leq n; i\leftarrow i+1)$ }
     {
     \STATE $a_i \leftarrow a_i - a_1$
     }
     \ENDFOR
   }
   \ENDWHILE
   \RETURN $a_1*2^p$
\end{algorithmic}
\end{algorithm}

Example 2: Let $a_1=14, a_2=28, a_3=56, a_4=98$ then \\
$GCD(a_1, a_2, a_3, a_4)$\\
$\Rightarrow GCD(14, 28, 56, 98)$\\
$\Rightarrow 2.GCD(7, 14, 28, 49)$\\
$\Rightarrow 2.GCD(7, 7, 14, 49)$\\
$\Rightarrow 2.GCD(7, 7, 7, 49)$\\
$\Rightarrow 2.GCD(7, 0, 0, 42)$\\
$\Rightarrow 2.GCD(7, 0, 0, 21)$\\
$\Rightarrow 2.GCD(7, 0, 0, 14)$\\
$\Rightarrow 2.GCD(7, 0, 0, 7)$\\
$\Rightarrow 2.GCD(7, 0, 0, 0)$\\
$\Rightarrow 2.7=14 $


\section{Conclusion and Future Work}
In this paper we have presented the algorithms for GCD computation of many integers. Traditional method of computing GCD of many integers by recursively calling Euclid's for each pair of integers can be expensive if the list of integers are not selected randomly. Future work can be to extend these algorithms to GCD computation of $n$ polynomials.

%
%

\end{document}